\documentclass[]{mn2e}

\usepackage{epsf}
\usepackage[dvips]{epsfig}
\usepackage{times}

\begin{document}

\title[Formation Scenario of W3]{A Possible Formation Scenario for the
  Ultra-Massive Cluster W3 in NGC~7252}

\author[M. Fellhauer, P. Kroupa]{M. Fellhauer, P. Kroupa \\
Sternwarte University Bonn, Germany}

\pubyear{2004}

\maketitle

\begin{abstract}
  The intermediate age star cluster W3 (age $\approx 300$--$500$~Myr)
  in NGC~7252 is the most luminous star cluster known to date with a
  dynamical mass estimate of $8 \pm 2 \cdot 10^{7}$~M$_{\odot}$.  With
  an effective radius of about $17.5$~pc and a velocity dispersion of 
  $45$~kms$^{-1}$ this object may be viewed as one of the recently
  discovered ultra-compact dwarf galaxies (UCDs).  Its intermediate
  age, however, precludes an origin as a stripped nucleated dwarf
  elliptical galaxy.
  
  The galaxy NGC~7252 is a merger remnant from two gas-rich disc
  galaxies.  Interactions between two gas-rich galaxies lead to bursts
  of intense star formation.  The age of the interaction and the age
  of W3 are in good agreement, suggesting that was W3 probably formed
  in the starburst.

  We propose a formation scenario for W3.  Observations of interacting
  galaxies reveal regions of strong star formation forming dozens up
  to hundreds of star cluster in confined regions of up to several
  hundred parsec in diameter.  The total mass of new stars in these
  regions can reach $10^{7}$ or even $10^{8}$~M$_{\odot}$.  By means
  of numerical simulations we have shown that the star clusters in
  these regions merge on short time-scales of a few Myr up to a few
  hundred Myr.  We apply this scenario to W3 and predict properties
  which could be observable and the future evolution of this object. 

  This work lends credence to the notion that at least some of the
  UCDs may be evolved star cluster complexes formed during early
  hierarchical mergers.
\end{abstract}

\begin{keywords}
  galaxies: NGC~7252 -- galaxies: interactions -- galaxies: formation
  -- galaxies: star clusters -- galaxies: dwarfs -- methods: N-body
  simulations 
\end{keywords}

\section{Introduction}
\label{sec:intro}

In the merger remnant system NGC~7252 Maraston et al.\ (2004) found
the most luminous star cluster known to date, W3. They determined its
mass via two independent ways.  First they used the total luminosity
of $M_{V} = -16.2$ and a stellar $M/L$ derived from a single stellar
population model to estimate its mass to be $7.2 \cdot
10^{7}$~M$_{\odot}$.  As an independent approach they calculated the
dynamical mass by measuring the velocity dispersion, which turned out
to be quite high, $\sigma = 45 \pm 5$~kms$^{-1}$, and the effective
radius of W3 $R_{\rm eff} = 17.5 \pm 1.8$~pc.  This translates to a
mass of W3 of $M_{\rm dyn} = 8 \pm 2 \cdot  10^{7}$~M$_{\odot}$
assuming dynamical equilibrium of a spherical object with isotropic
velocity distribution.  The age of this object is about
$300$--$500$~Myr (Maraston et al.\ 2004 and references therein), which
indicates that it probably formed during the merger event of the
host system.  The projected distance of W3 to the centre of NGC~7252
is about $10$~kpc and W3 appears to lie within the optical
radius of the galaxy.  The size and mass of this object leads to the
suggestion that it may be one of the recently discovered ultra compact
dwarf galaxies (UCD) found in the Fornax cluster (Hilker et al.\ 1999;
Phillipps et al.\ 2000), rather than an 'ordinary' globular cluster. 
The UCD galaxies have been suggested to be the cores of stripped
nucleated dwarf galaxies (Bekki et al.\ 2003, Mieske et al.\ 2004).

We propose instead a formation scenario which is closely related to
the massive star-bursts caused by the interaction of two gas-rich disc
galaxies.  In interacting systems like the Antennae (NGC~4038/39;
Whitmore et al.\ 1999, Zhang \& Fall 1999) regions of very intense
star-formation arise as a result of the tightly compressed
interstellar media.  Dozens and up to hundreds of young massive star
clusters are observed to form in star cluster complexes (or {\it
  super-clusters}) spanning up to a few hundred pc in diameter.
Kroupa (1998) argues that these super-clusters have to be bound
objects because their age ($\approx 10$~Myr) indicates that they
should be already dispersed.  Simulating super-clusters, by means of
stellar dynamical $N$-body simulations, Fellhauer et al.\ (2002) found
that the star clusters within these super-clusters merge on very short
time-scales (a few dozens to a few hundred Myr), namely a few
crossing-times of the super-cluster.

The resulting merger-objects can be characterised by their large
effective radii.  They show similar properties like the UCDs
(Fellhauer \& Kroupa 2002a) or the faint fuzzy star clusters
(Fellhauer \& Kroupa 2002b) found in a survey of S0 galaxies (Larsen
\& Brodie 2000), that are understood to be morphological types
resulting from the merger of at least two late-type galaxies. 

To distinguish between the two formation scenarios for UCD galaxies
one would need detailed information about the metallicities and ages of 
the stellar populations inside these objects.  Dwarf galaxies have a
rather complex mixture of stellar populations, while star clusters
form in a single mono-metallic starburst.  Unfortunately this
information is not yet accessible for objects as distant as the UCDs
in Fornax or W3 in NGC~7252.  

However, as pointed out by Kroupa (1998), objects as massive as the
super clusters firstly can retain stellar winds thereby progressively
building up an interstellar medium from which new stars may later
form, secondly can capture old field stars during their formation and
thirdly accrete gas later-on leading to younger stellar generations.
The stellar age and metalicity distribution may thus be very complex
in evolved super-clusters. 

The age-estimate of W3 points to the possibility that it may have
formed during the merger event of the host-galaxy (i.e.\ out of
merging star clusters) rather than it being the stripped core of a
dwarf galaxy, which should be very old.

Another indication of the nature of this object would come from
a detailed surface-brightness profile.  Young massive star clusters
formed in the Large Magellanic Cloud in strong star burst regions
show ripples in their profiles (Schweizer 2004).  This also suggests
a formation history through the merging of several objects. 

In this project we perform stellar dynamical N-body simulations to
show that the merging of star clusters in dense star cluster complexes
is able to form massive objects like W3 in NGC~7252. In the next
section we describe the setup of our models followed by the results of
our investigation.  Finally we end with a discussion of our results.

\section{Setup}
\label{sec:setup}

The simulations are carried out with the particle-mesh code {\sc
  Superbox} with high-resolution sub-grids which stay focussed on the
simulated objects (Fellhauer et al.\ 2000).  

In our models the super-cluster is initially represented by a Plummer
sphere with a Plummer radius of $100$~pc and a cut-off radius of
$500$~pc.  The clusters inside this super-cluster have a total mass of
$9.9 \cdot 10^{7}$~M$_{\odot}$, which leads to a crossing time of the
super-cluster of $t_{\rm cr,sc}9.3$~Myr.  Inside the Plummer sphere of
the super-cluster, the 'particles' have positions and velocities
according to the Plummer distribution function.  The 'particles'
themselves are Plummer spheres representing the star clusters with,
respectively, Plummer radii of $4$ or $10$~pc, masses of $10^{6}$ and
$5 \cdot 10^{6}$~M$_{\odot}$ and crossing times of $0.75$ and
$1.32$~Myr, being represented by $10^{5}$ and $5 \cdot
10^{5}$~particles.  Altogether the super-cluster is filled with $75$
star clusters, from which $69$ are light ones and $6$ are of the heavy
type to mimic a power law mass spectrum similar (power law index
$\alpha = -1.9$) to the one found in the young massive star clusters
in the Antennae (Zhang \& Fall 1999).   
To model hundreds of lighter and maybe thousands of ultra-light star
clusters is beyond the abilities of our code and so we placed more
mass into massive clusters.  Our previous work showed that applying a
mass spectrum to the clusters does not affect the results, because the
first build up of a heavy merger object in the centre of the
super-cluster happens on a time-scale much shorter than the time-scale
of mass-segregation or relaxation (i.e.\ shorter than $t_{\rm
  cr,sc}$).  The effective radius of the light clusters is chosen
according to the mean effective radius of the star clusters in the
Antennae (Whitmore et al.\ 1999) while the effective radius of the
heavy ones represents merger-objects of many light star 
clusters, so that the overall model is such that progressively more
massive clusters are constructed hierarchically from star clusters
(i.e.\ truly mono-metalicity and mono-age populations) with a finite
upper mass limit (Weidner \& Kroupa 2004).  Such a hierarchical model
imposes no physical limit on the mass of the composite systems, and in
principle entire galaxies can be viewed as being such objects.  This
also means that our model does not start at $t=0$ of the
formation but after a time $t<t_{\rm cr,sc}$ when the super-cluster is
already free of its residual gas, as seen in the Knots of the
Antennae, which are surrounded by $H_{\alpha}$ bubbles (Whitmore et
al.\ 1999). 

Initially the single star clusters and the whole star cluster complex
are in virial equilibrium, being set-up from the Plummer distribution
function (Aarseth et al. 1974).  This corresponds to a high
star-formation efficiency for such starburst objects (Elmegreen \&
Efremov 1997).

This super-cluster is now placed into an analytical galactic potential
consisting of a logarithmic potential for the underlying halo, a
Plummer-Kuzmin disc and a Hernquist bulge, which add up to an almost
flat rotation curve of $220$~kms$^{-1}$.  The super-cluster is placed
at a distance of $20$~kpc initially (apogalacticon) on an eccentric
orbit with perigalacticon of $10$~kpc.

The grid-levels of {\sc Superbox} are chosen in a way that the
innermost grids with the highest resolution cover the single star
clusters resolving intra-cluster forces and the forces during the
merging of two star clusters.  The median resolution grid is chosen to
cover the area of the whole super-cluster, resolving the forces
between the clusters.  Finally the outermost, non-moving grid covers
the orbit of the super-cluster around the host galaxy.  A detailed
listing of the grid-parameters can be found in Tab.~\ref{tab:grid}.

\begin{table}
  \centering
  \caption{Grid sizes (size) and resolutions (resol.) of the different
    grid-levels of the objects in the {\sc Superbox} simulation.} 
  \label{tab:grid}
  \begin{tabular}[t!]{lrrrrrr} \hline
    Object & \multicolumn{2}{c}{inner grid} &
    \multicolumn{2}{c}{medium grid} & \multicolumn{2}{c}{outer grid}
    \\ 
    & size & resol. & size & resol. & size & resol.
    \\
    & [pc] & [pc] & [pc] & [pc] & [kpc] & [kpc] \\ \hline
    light cluster & 50.0 & 0.83 & 1000.0 & 16.67 & 50.0 & 0.83 \\
    heavy cluster & 120.0 & 2.00 & 1000.0 & 16.67 & 50.0 & 0.83 \\
    merger object & 120.0 & 2.00 & 1000.0 & 16.67 & 50.0 & 0.83 \\
    \hline 
  \end{tabular}
\end{table}

\section{Results}
\label{sec:res}

In contrast to our previous models (Fellhauer \& Kroupa 2002a,b) where
the merger objects were smaller and not that heavy, the merging
process here takes more time.  After the first clusters have merged
and build up a massive and extended object with the remaining clusters
within, it is not encounters between clusters and the merger object
which govern the merging process (the clusters are already within the
object) but dynamical friction acting on the remaining clusters.
Nevertheless a massive merger object is present from an early stage on.

\begin{figure}
  \centering
  \epsfxsize=03.9cm
  \epsfysize=04.0cm
%  \epsffile{w02-t300.eps}
  \epsffile{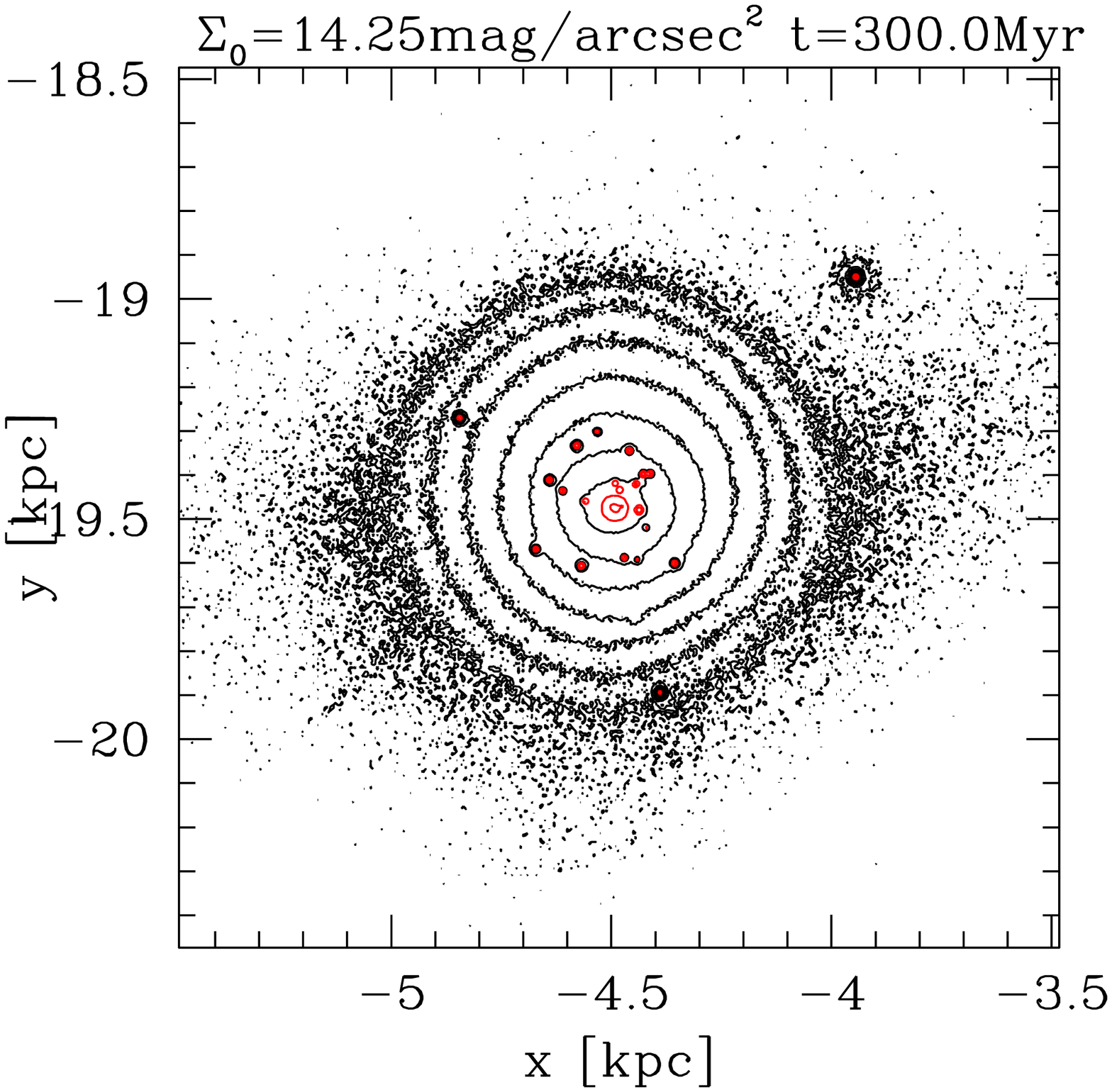}
  \epsfxsize=03.9cm
  \epsfysize=04.0cm
%  \epsffile{w02-t500.eps}
  \epsffile{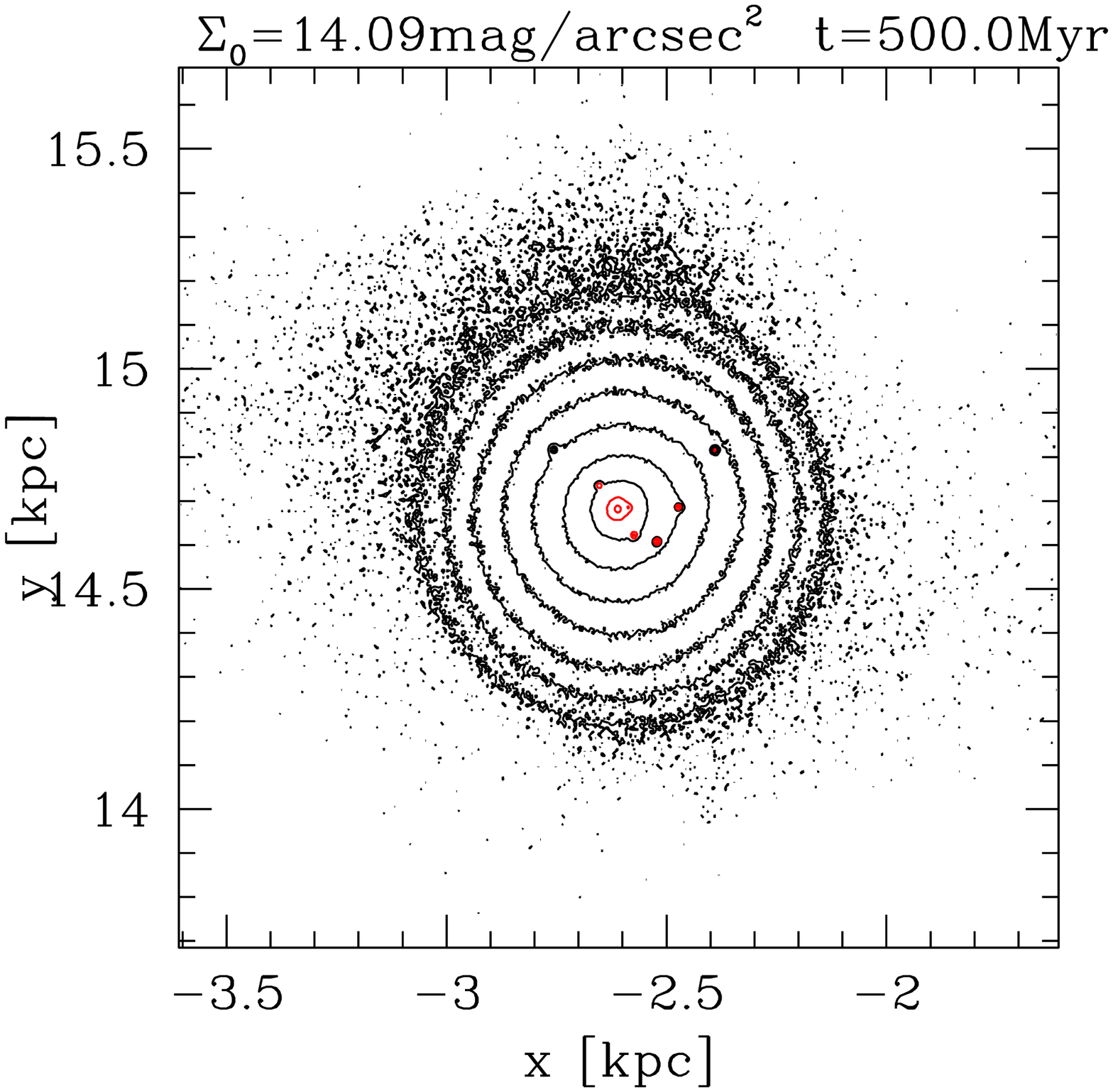}
  \epsfxsize=04.0cm
  \epsfysize=04.0cm
%  \epsffile{w02t300-surf.eps}
  \epsffile{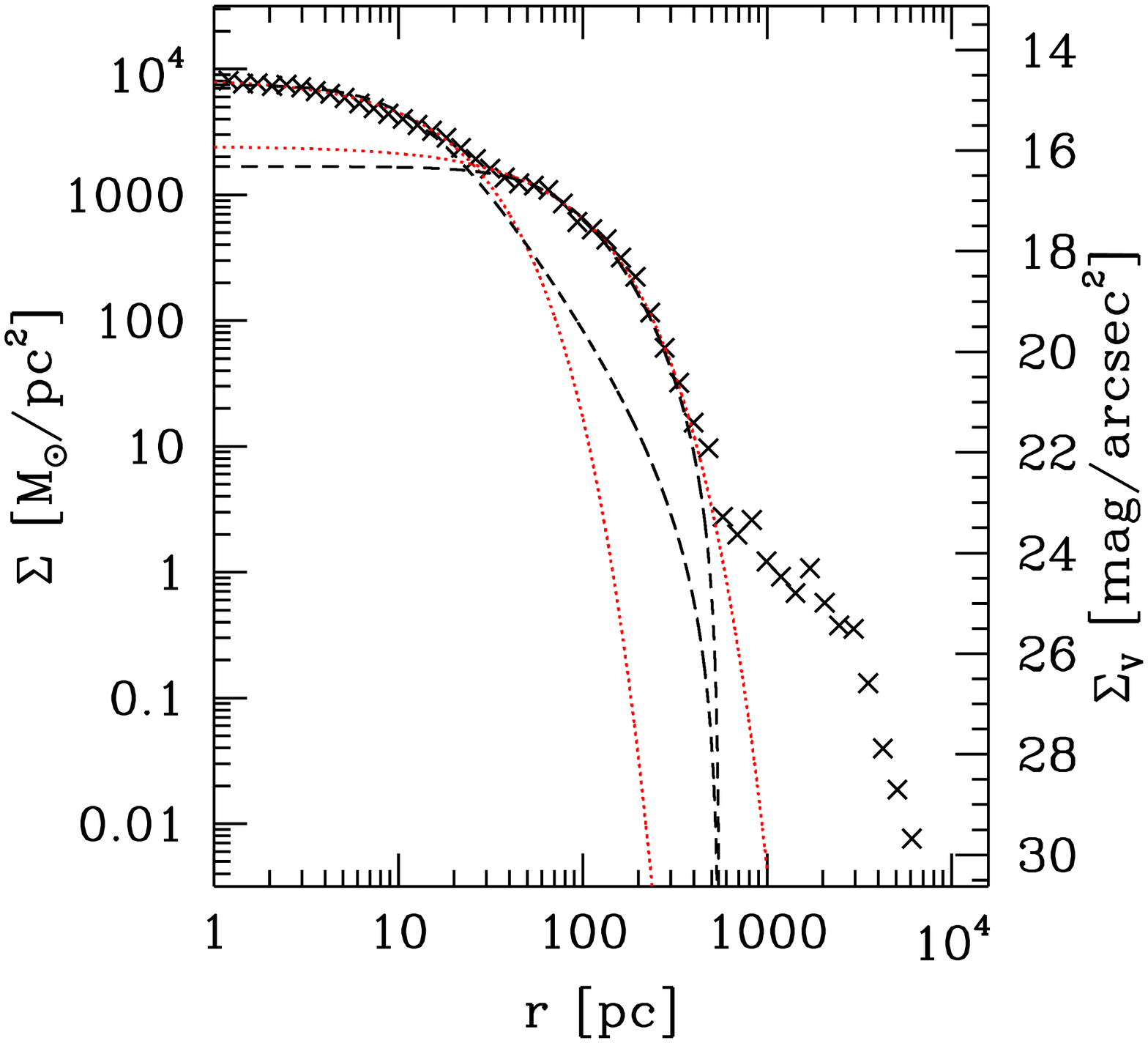}
  \epsfxsize=04.0cm
  \epsfysize=04.0cm
%  \epsffile{w02t500-surf.eps}
  \epsffile{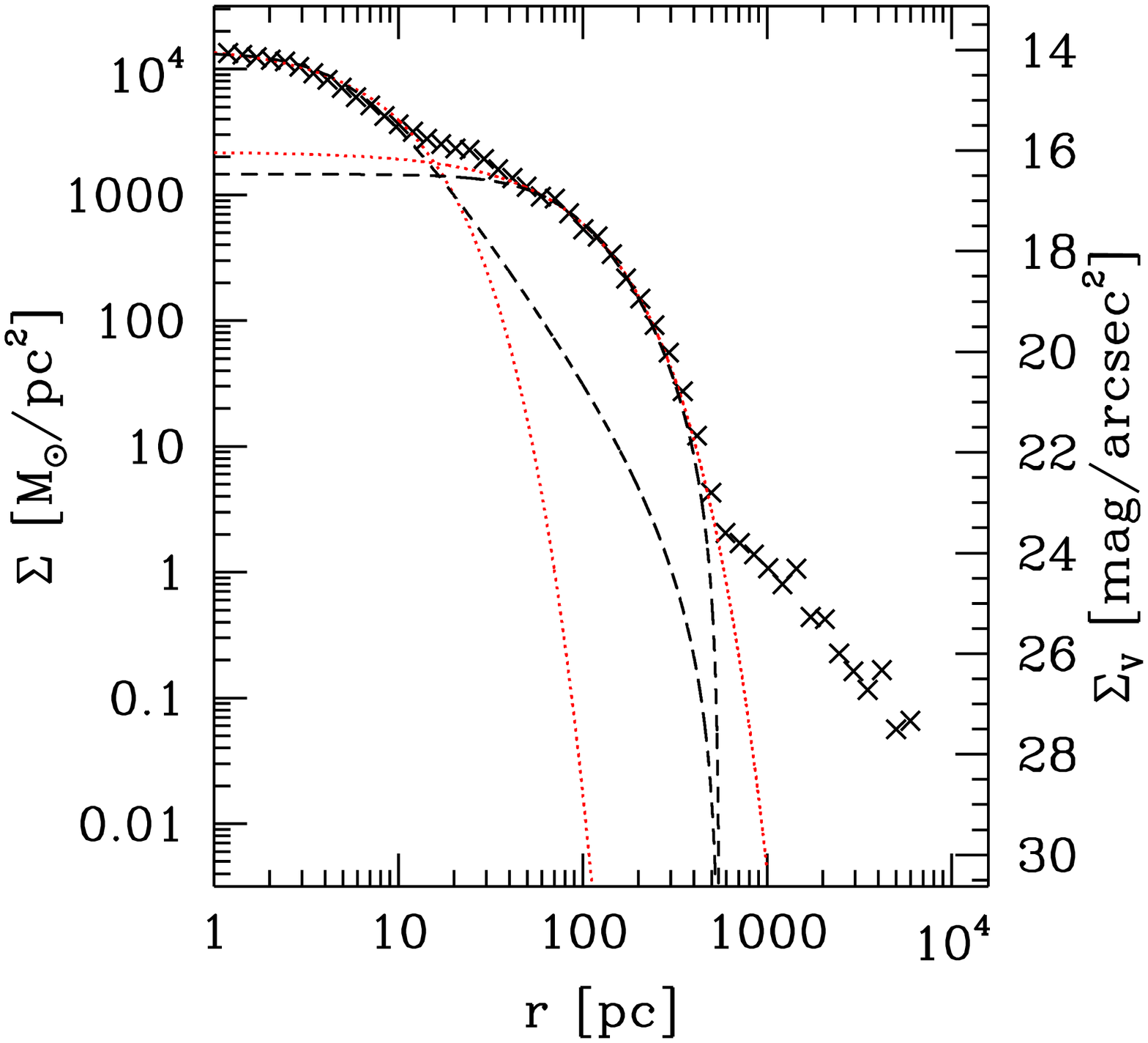}
  \epsfxsize=04.0cm
  \epsfysize=04.0cm
%  \epsffile{w02t300-vlos.eps}
  \epsffile{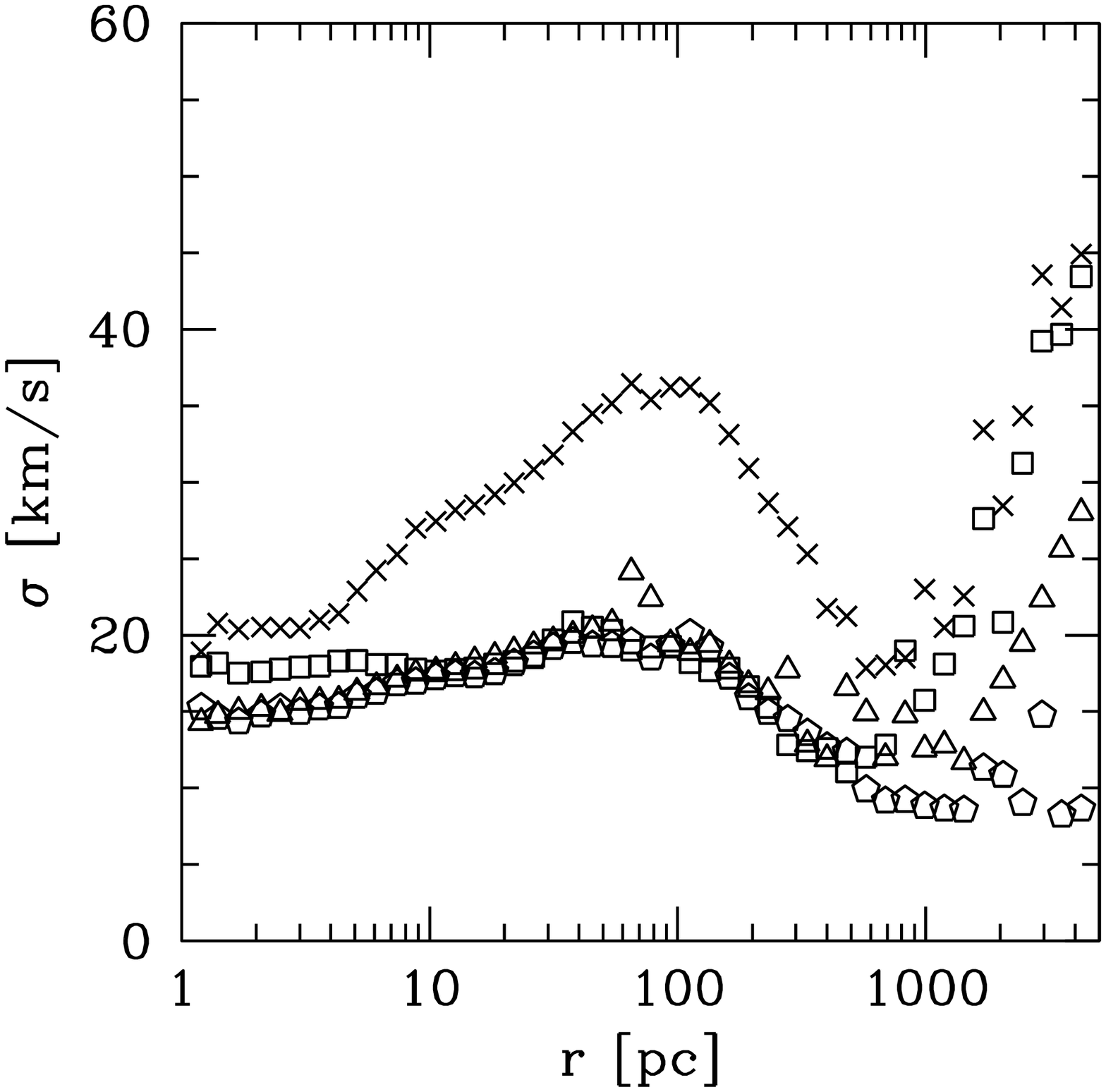}
  \epsfxsize=04.0cm
  \epsfysize=04.0cm
%  \epsffile{w02t500-vlos.eps}
  \epsffile{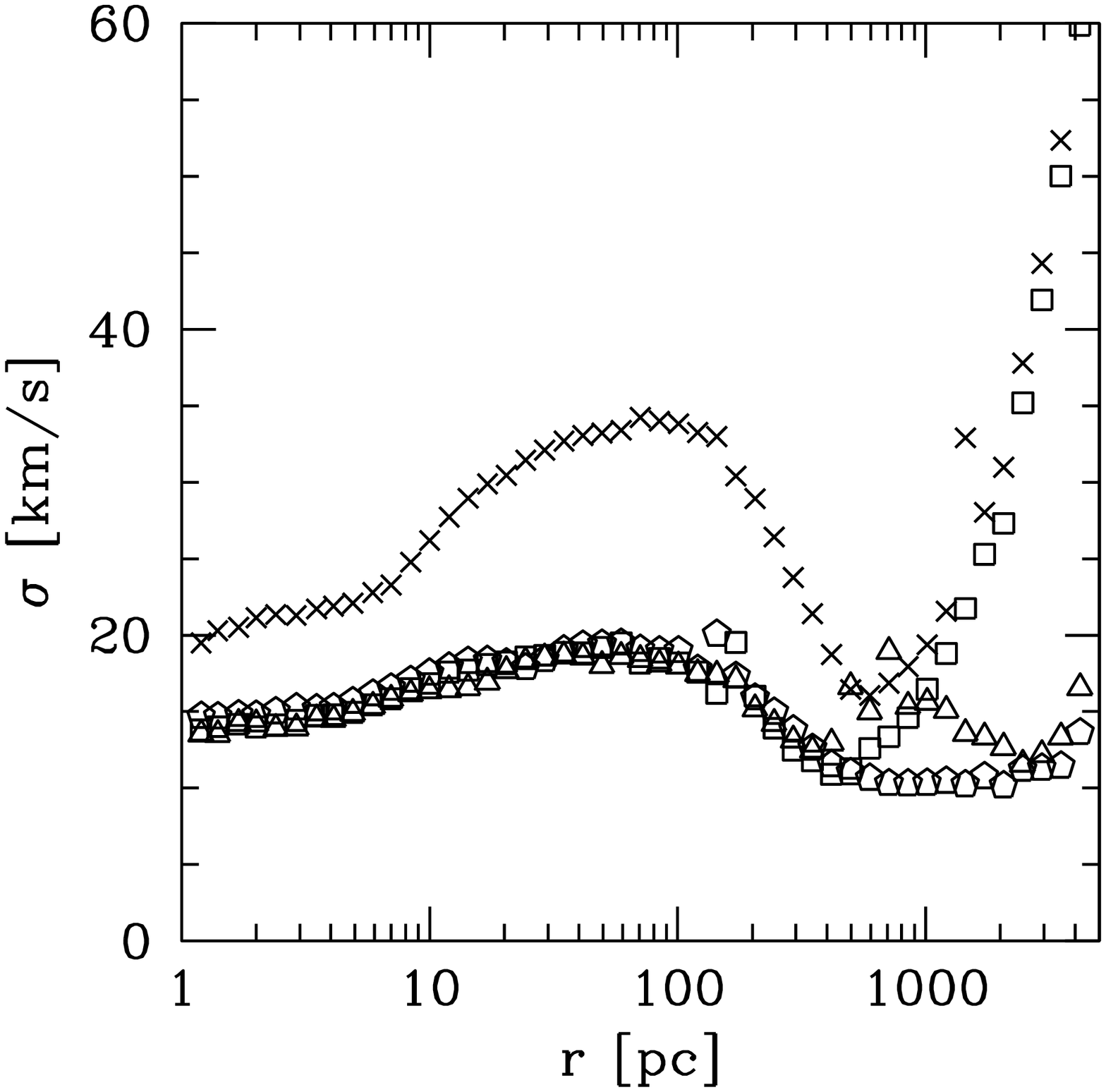}
  \caption{Left: W3 model at $t=300$~Myr; Right: W3 model at
  $t=500$~Myr.  Top: Contour plot taking a $M/L=0.15$.  Contours have
  magnitude spacings; outermost contour corresponds to a surface
  brightness of $23$~mag.arcsec$^{-2}$.  Middle: Surface density
  profile of the merger object.  Crosses are the data points, dashed
  lines are the fit with two King profiles, dotted lines are the fit
  with two exponentials.  Brightnesses on the right are calculated
  using $M/L=0.15$.  Bottom: Velocity dispersions; open symbols show
  the line-of-sight dispersions measured in concentric rings around
  the centre of density of the object along the coordinate axes.
  Crosses are the 3D velocity dispersion measured in concentric
  shells.}  
  \label{fig:t300}
\end{figure}

The top row of Figure~\ref{fig:t300} shows the contour plot of the
merger object (left at $t=300$ and right at $t=500$~Myr comprising the
suggested age-range of W3).  The contours are spaced in magnitude
intervals and masses are converted to luminosities taking a
mass-to-light ratio of 0.15.  According to a single stellar population
model run with Starburst99 (Leitherer et al.\ 1999) a stellar population 
of this age has a $M/L=0.1$--$0.2$.  In our model these two
time-slices mark the transition between the time, when still many star
clusters are visible as separate entities and the time when most of
the star clusters have already merged or are dissolved within the
merger object.  The unmerged star clusters also cause little wiggles
in the surface density profile.  Unfortunately NGC~7252 and thus W3
are too far away to resolve the surface density profile in detail.
Wiggles like this are found in resolved young massive star clusters
in the LMC (Schweizer 2004). 

\begin{table}
  \centering
  \caption{The fitting parameters for the W3 model using King or
    exponential functions.} 
  \label{tab:fit}
  \begin{tabular}[t!]{r|r|r|r} \hline
     & $t=300$~Myr & $t=500$~Myr & $t=5$~Gyr \\ \hline
     \multicolumn{4}{l}{\it King Surface Density Profiles} \\
     tidal radius [pc] & $550$ & $550$ & $550$ \\
     {\bf core:} & & & \\
     core radius [pc] & $12.4 \pm 0.8$ & $5.8 \pm 0.4$ & $11.3 \pm
     0.4$ \\
     $\Sigma_{0}$ [M$_{\odot}$/pc$^{2}$] & $7930 \pm 190$ & 
     $13820 \pm 450$ & $7240 \pm 110$ \\
     {\bf envelope:} & & & \\
     core radius [pc] & $94.2 \pm 6.4$ & $98.5 \pm 5.5$ & $84.5 \pm
     1.3$ \\
     $\Sigma_{0}$ [M$_{\odot}$/pc$^{2}$] & $2430 \pm 90$ & $2150 \pm
     60$ & $1510 \pm 20$ \\ \hline
     \multicolumn{4}{l}{\it Exponential Surface Density Profiles}\\
     {\bf core:} & & & \\
     exp.\ radius [pc] & $16.1 \pm 0.8$ & $7.3 \pm 0.5$ & $14.7 \pm
     0.5$ \\ 
     $\Sigma_{0}$ [M$_{\odot}$/pc$^{2}$] & $8410 \pm 160$ & $15460 \pm
     600$ & $7730 \pm 100$ \\
     {\bf envelope:} & & & \\
     exp.\ radius [pc] & $75.6 \pm 3.6$ & $76.1 \pm 3.2$ & $75.4 \pm
     0.6$ \\ 
     $\Sigma_{0}$ [M$_{\odot}$/pc$^{2}$] & $2430 \pm 130$ & $2190 \pm
     110$ & $1680 \pm 20$ \\ \hline
     \multicolumn{4}{l}{\it Exponential 3D Velocity Dispersion
       Profile (\rm outer part)} \\
     exp.\ radius [pc] & $591 \pm 23$ & $515 \pm 13$ & $435 \pm 7$ \\
     $\sigma_{0}$ [km/s] & $43.3 \pm 0.7$ & $41.2 \pm 0.5$ & $39.4 \pm
     0.3$ \\ \hline
  \end{tabular}
\end{table}

Another interesting feature of our model is that the merger object
shows a cuspy structure with a dynamically cold core.  The profile can 
be fitted by either two King profiles or two exponentials
(Fig.~\ref{fig:t300}, middle row) and the velocity dispersion is
rising in the innermost part and reaches its maximum beyond the
transition between the core and the envelope (Fig.~\ref{fig:t300},
bottom row).  The outer part of the velocity dispersion, reaching from
the maximum value to the tidal radius, can be fitted with an
exponential profile with an exponential scale length of the order of
the tidal radius of the object.  A detailed listing of the fitting
parameters can be found in Tab.~\ref{tab:fit}.   Beyond the tidal
radius the velocity dispersion is rising again due to the large
velocities of the extra tidal stars. 

\begin{figure}
  \centering
  \epsfxsize=04.0cm
  \epsfysize=04.0cm
%  \epsffile{w02-mass.eps}
  \epsffile{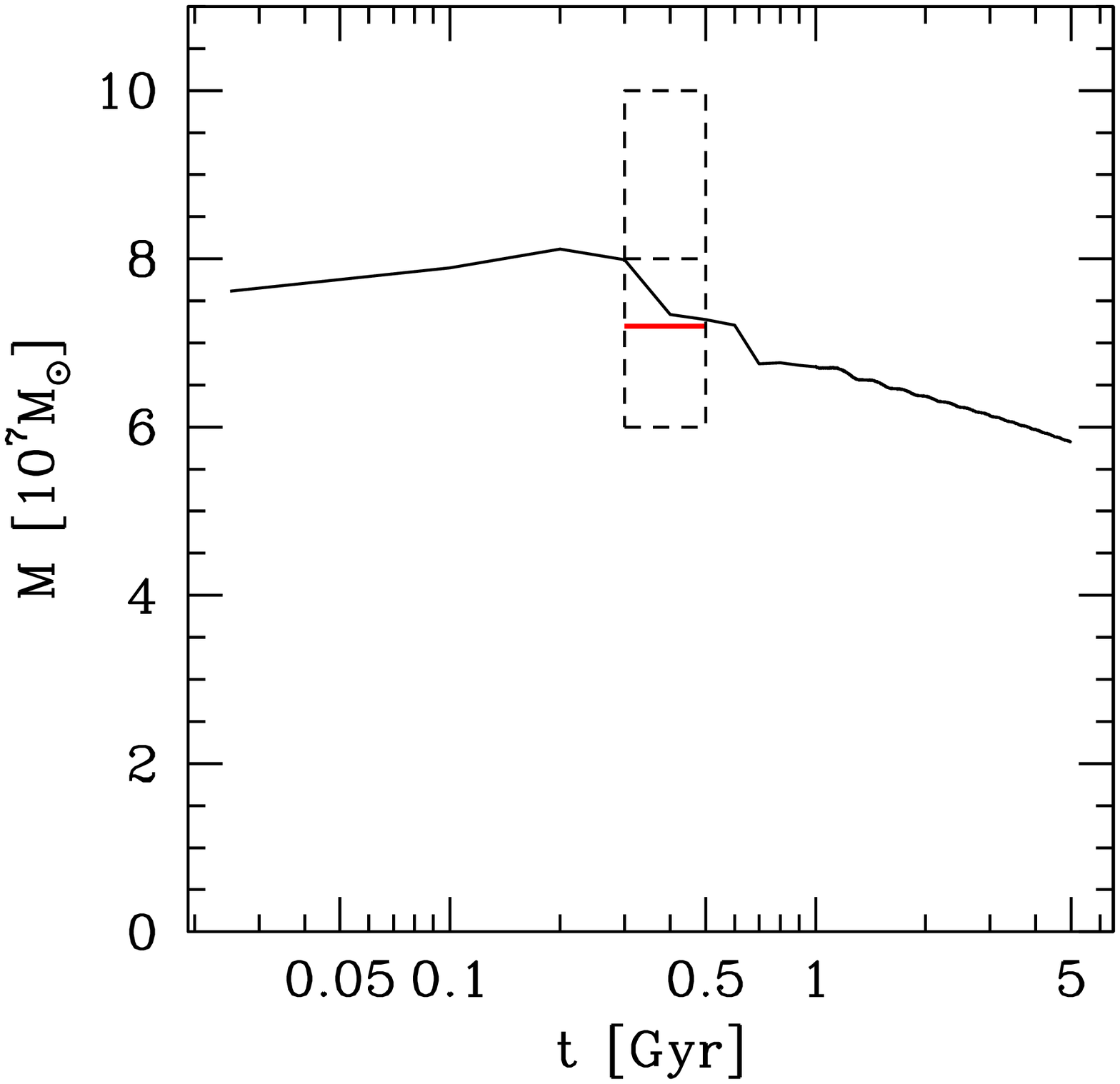}
  \epsfxsize=04.0cm
  \epsfysize=04.0cm
%  \epsffile{w02-reff.eps}
  \epsffile{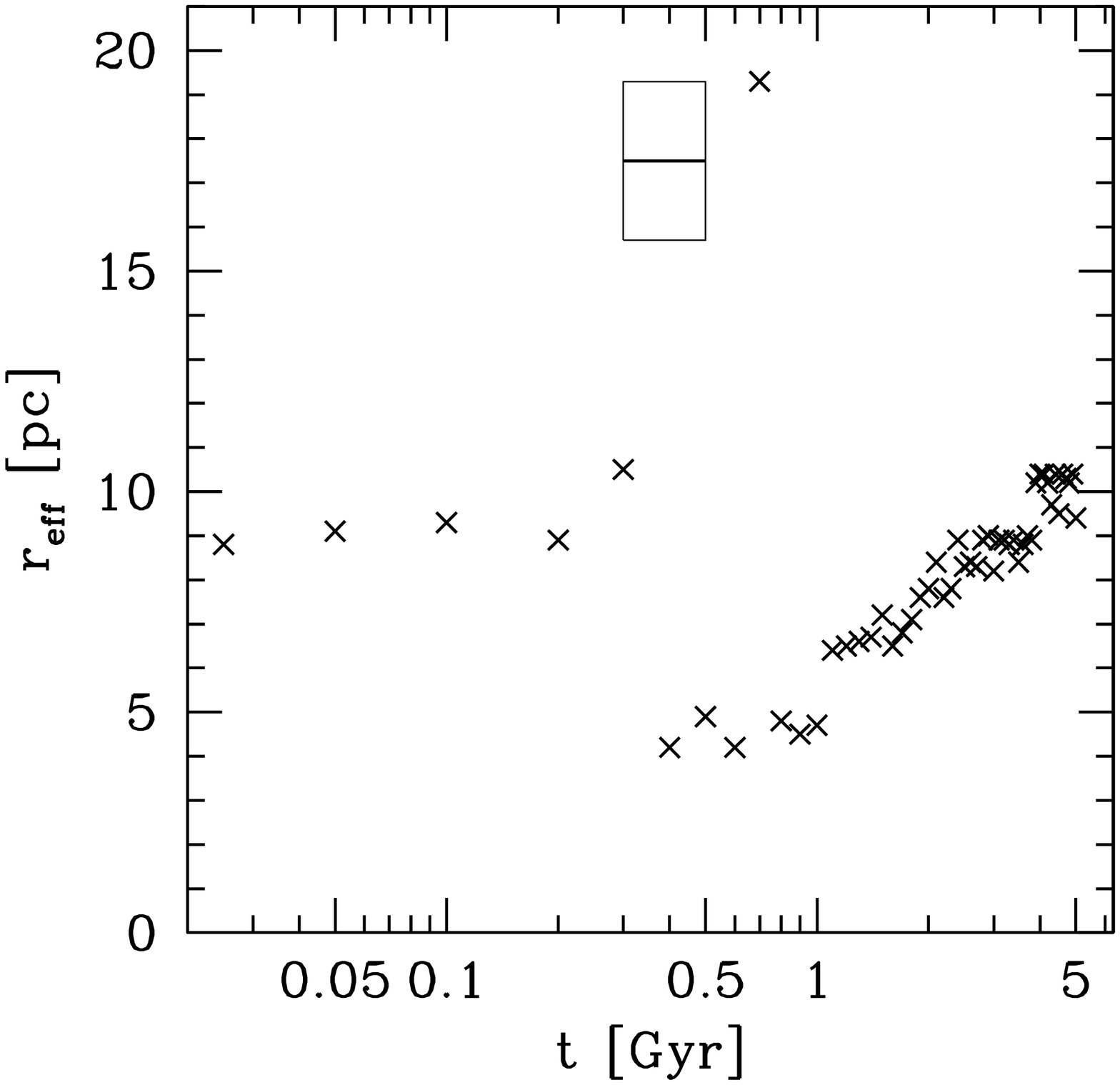}
  \epsfxsize=04.0cm
  \epsfysize=04.0cm
%  \epsffile{w02-sigm.eps}
  \epsffile{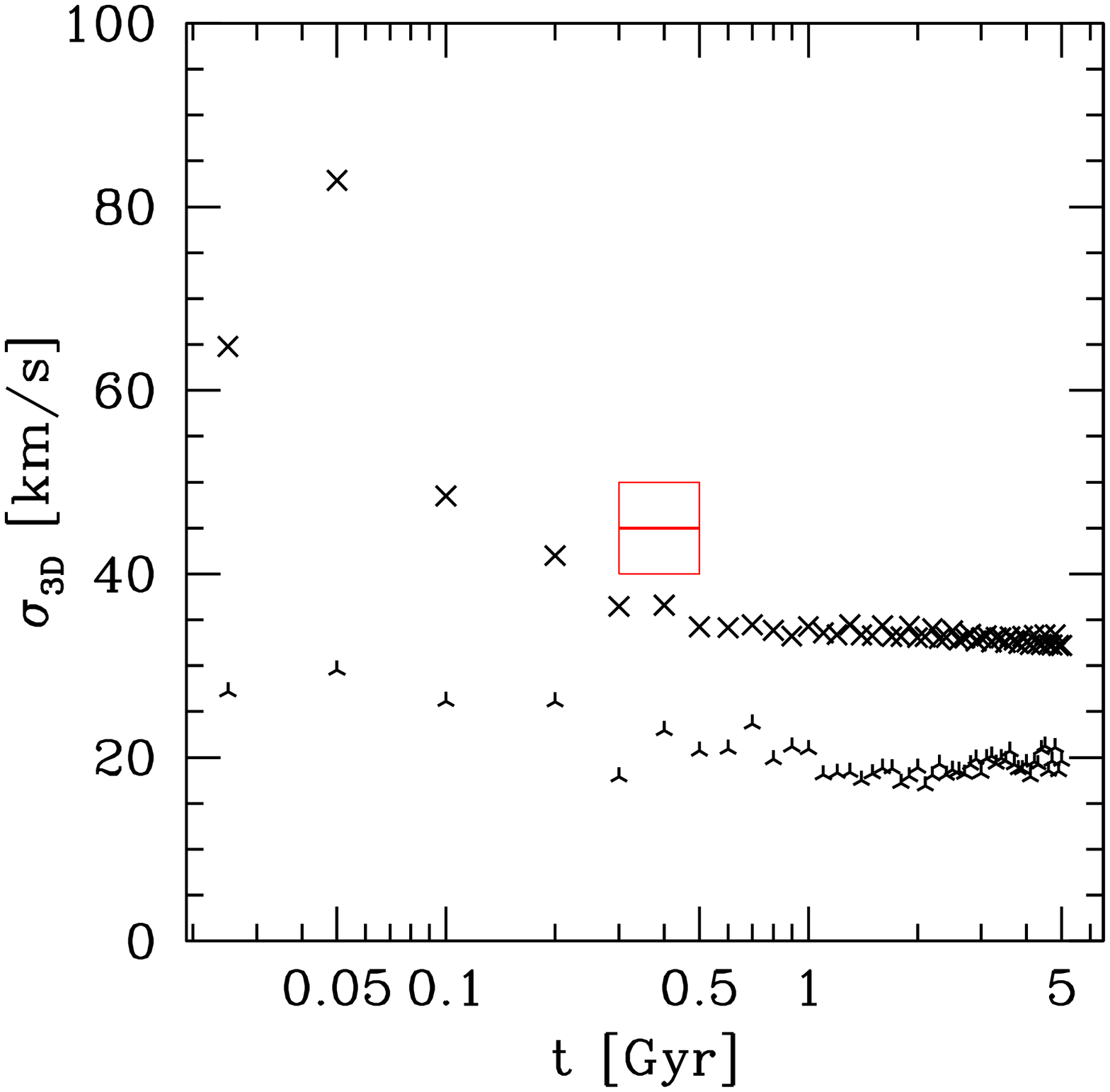}
  \caption{Time evolution of characteristic parameters of our model.
    Total mass (top left panel), thick horizontal line shows the value
    derived from the luminosity of W3 dashed line and box shows the
    dynamical mass and the $1\sigma$-uncertainty of W3; Effective
    radius (top right panel), horizontal line shows the fitted value
    of W3 and the box around denotes the $1\sigma$-uncertainty;
    Velocity dispersion (bottom panel): Crosses are the maximum
    velocity dispersion values and three-pointed stars are the central
    values.  Horizontal line and box again shows the measured value
    for W3 with $1\sigma$-uncertainty.} 
  \label{fig:evol}
\end{figure}

In Fig.~\ref{fig:evol} we show the evolution of the total mass, of the 
effective radius and of the velocity dispersion of our model, whose
dynamical evolution was followed for $5$~Gyr.  To derive the mass we
only take the bound particles into account.  The effective radius is
taken at the point where the surface brightness has dropped to half of
the central value. Finally the 3D-velocity dispersion was derived in
concentric shells around the centre of the object.  Shown in the
figure are the central values of the dynamically cold core as well as
the maximum values.  The boxes show the values measured for W3.
The total mass as well as the maximum velocity dispersion of our model
is in reasonable agreement with the data of W3.  Only the effective
radius of our model is too small compared with W3.  It starts off at a
value of about $10$~pc and it increases slightly until the majority of
the star clusters have merged, deforming the core progressively.
After the merging process has ceased the core of the merger object
stabilises and the effective radius drops to about $5$~pc.  But note
the extremely large value at $t=700$~Myr.  At exactly that time a late
merger event of one of the remaining star clusters happened.  If our
formation scenario is correct then this points to the possibility that
we might be observing W3 at the time when a star cluster merges with
the core of the object mimicking a large effective radius.  

The further dynamical evolution is mainly governed by the tidal
shaping of the object due to its eccentric orbit.  It leads to a
successive mass loss and an increase in the effective radius as well
as a decrease in the maximum velocity dispersion, while the central
value increases slightly. 

\begin{figure}
  \centering
  \epsfxsize=04.0cm
  \epsfysize=04.0cm
%  \epsffile{w02-t5g.eps}
  \epsffile{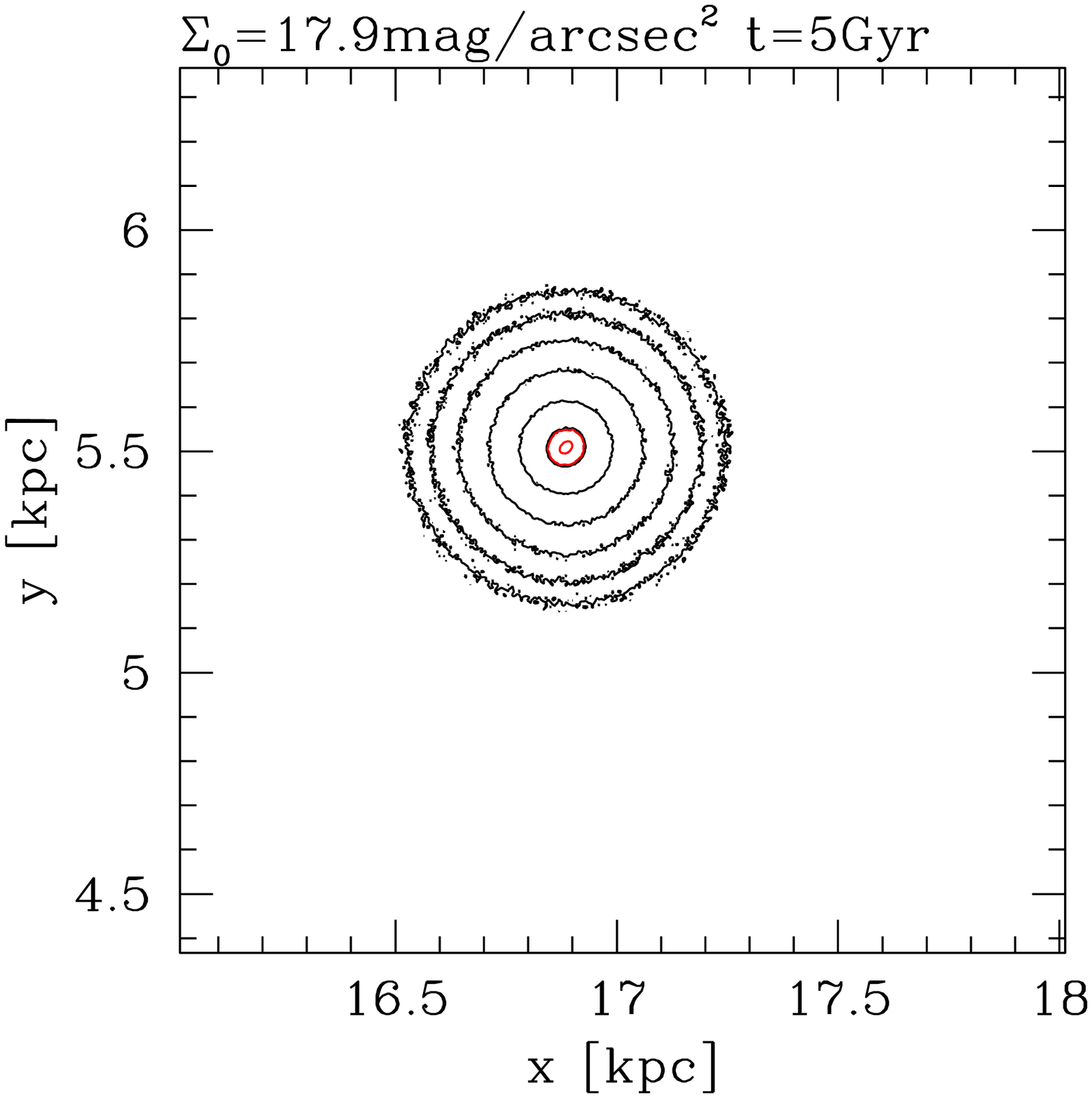}
  \epsfxsize=04.0cm
  \epsfysize=04.0cm
%  \epsffile{w02t5g-surf.eps}
  \epsffile{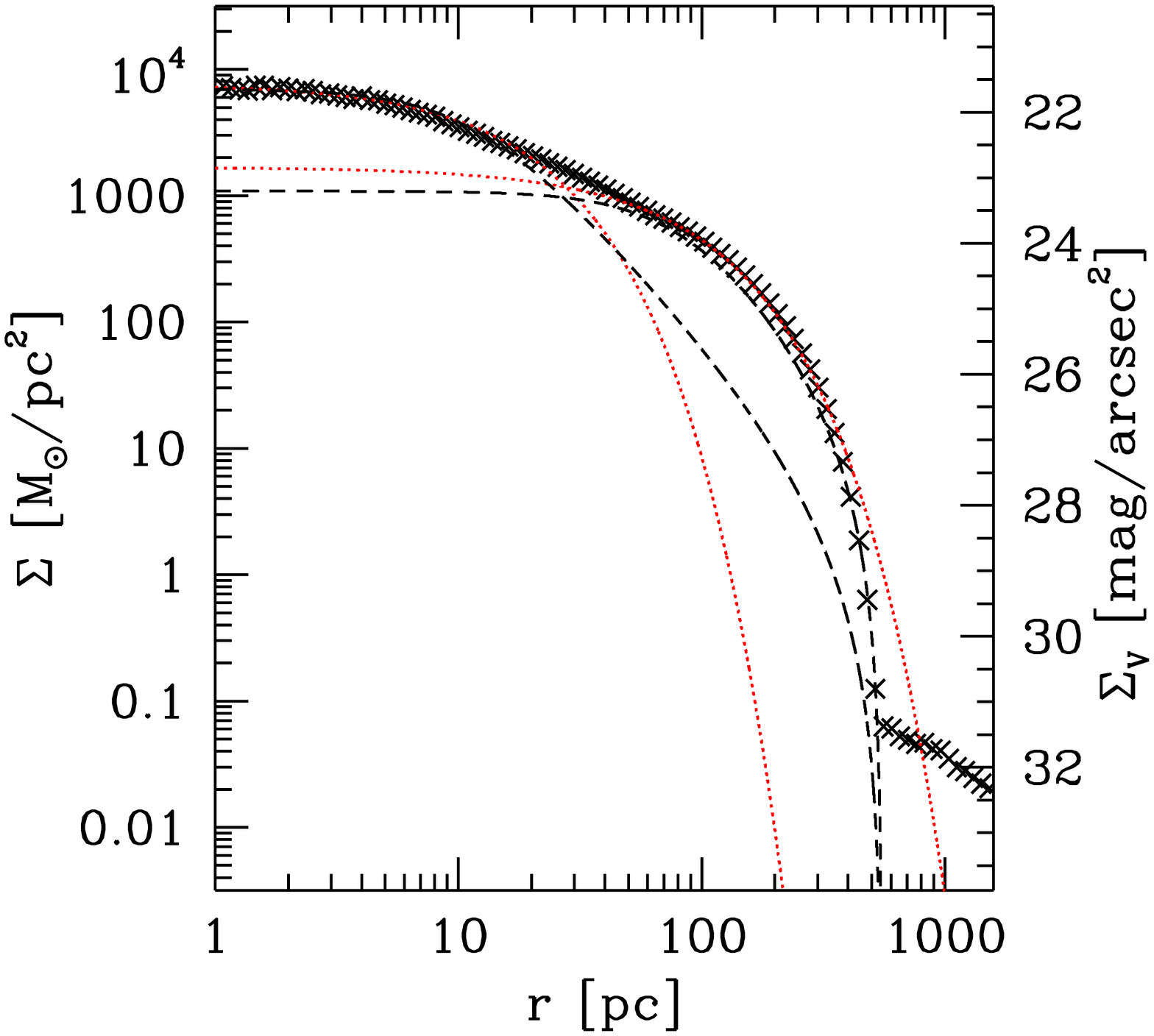}
  \epsfxsize=04.0cm
  \epsfysize=04.0cm
%  \epsffile{w02t5g-vlos.eps}
  \epsffile{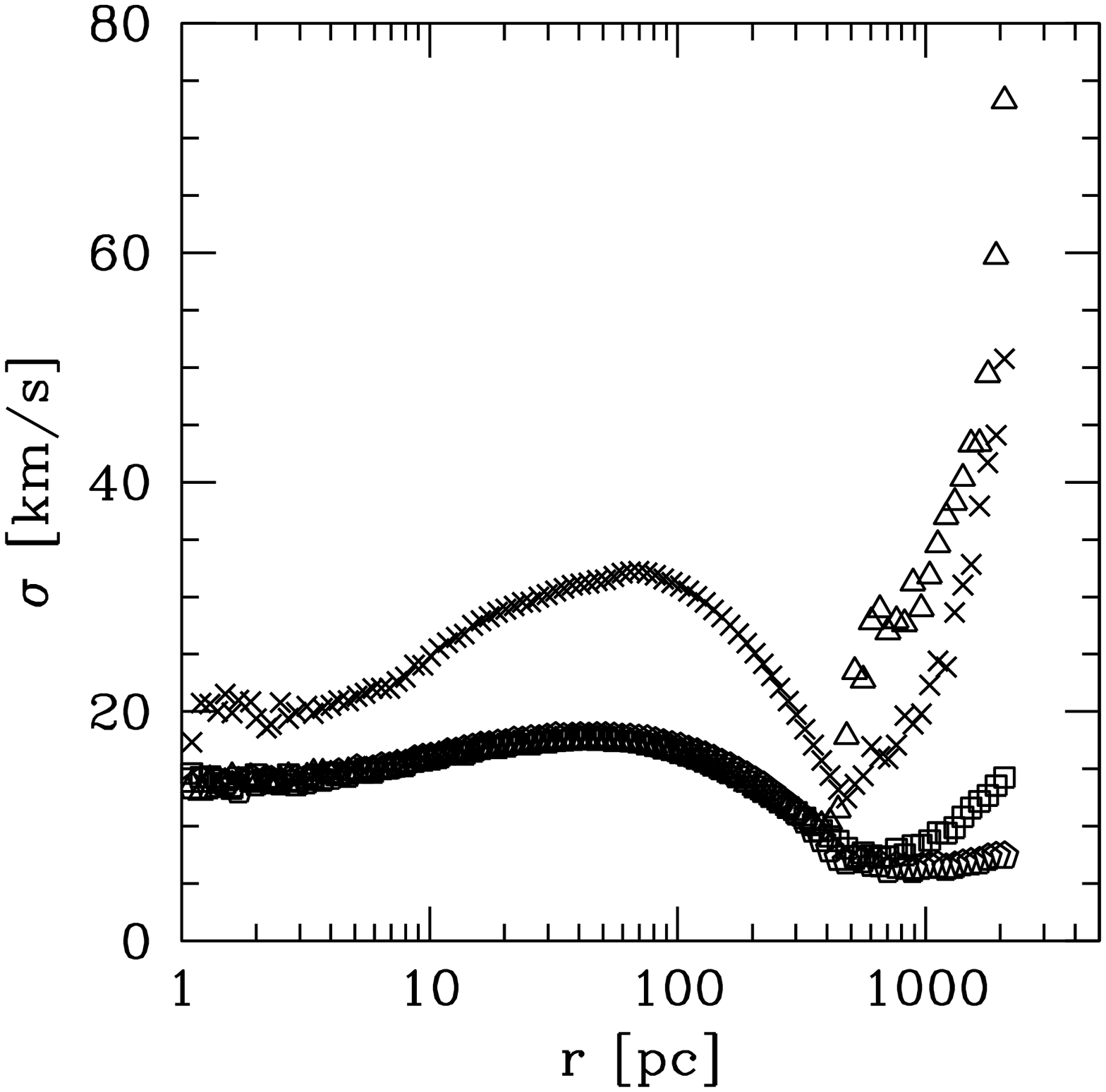}
  \caption{Merger Object at $5$~Gyr.  In the contour plot (outermost
    contour corresponds to $\Sigma_{0}=25$~mag.arcsec$^{-2}$) and the
    magnitude scale of the surface density profile $M/L=3.0$ is
    adopted.  The symbols and lines have the same meaning as in
    Fig.~\ref{fig:t300}}   
  \label{fig:t5gyr}
\end{figure}

Looking at the merger object after $5$~Gyr of its dynamical evolution
(Fig.~\ref{fig:t5gyr}) shows that this object is very stable against
tidal disruption.  Also the dynamically cold core is not a transient
feature but survives the evolution.  The total mass of the object is
still of the order of $6 \cdot 10^{7}$~M$_{\odot}$ which is an order
of magnitude more than the most massive globular cluster
($\omega$~Cen) of the Milky Way.  The relaxation time-scale of this
object, adopting the formula from Spitzer \& Hart (1971), is
\begin{eqnarray}
  \label{eq:relax}
  t_{\rm relax} & = & 0.138 \cdot \frac{\sqrt{M_{\rm cl}}
    r_{h}^{3/2}} {\left< m \right> \sqrt{G} \ln(0.4\,n)},
\end{eqnarray}
where $M_{\rm cl}$ is the mass of the object, $r_{h}$ is
the half-mass radius ($151.7$~pc at $t=5$~Gyr), the average stellar
mass $\left< m \right> = 0.4$~M$_{\odot}$ (using the universal IMF,
Kroupa 2001) and $n$ the number of stars.  This leads to $t_{\rm
  relax}\approx 4000$~Gyr.  This shows that this object might to be
thought of as a dwarf galaxy rather than a globular cluster, since
$t_{\rm relax}$ is much longer than a Hubble time.
 
\begin{figure}
  \centering
  \epsfxsize=08.0cm
  \epsfysize=08.0cm
%  \epsffile{w02-korm.eps}
  \epsffile{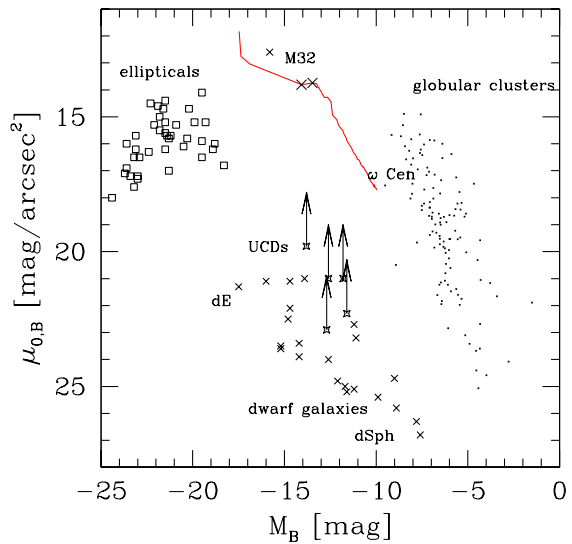}
  \caption{Central surface brightness, $\mu_{0,{\rm B}}$, vs absolute
      photometric B-band magnitude, $M_{\rm B}$ (the Kormendy
      diagram).  The small dots are data for Milky-Way globular
      clusters (Harris 1996), Local-Group dwarf galaxies (Mateo
      1998) are marked as crosses and elliptical galaxies (Peletier
      et al.\ 1990) are shown as open boxes.  The newly discovered
      UCDs are shown by arrows with  lower-limits on $\mu_{\rm B}$
      (Phillipps et al.\ 2001).  The line shows the evolution of our
      merger object from the time of formation until 5~Gyr using a
      time-dependent $M/L_{B}$ from a single stellar population model
      of Starburst99 (Leitherer et al.\ 1999). Crosses on the line
      mark the time between $300$ and $500$~Myr.} 
  \label{fig:korm}
\end{figure}

Finally we convert the evolution of the total mass and the central
surface density into total luminosity and central surface brightness
adopting the mass-to-light ratios from the evolution of a single
stellar population calculated with Starburst99 (Leitherer et al.\
1999).  This enables us to place the evolution of our model into a
Kormendy diagram (Fig.~\ref{fig:korm}).  It starts above the values
for the Knots (super-clusters) in the Antennae, because our
super-cluster model is much more massive than the super-clusters seen 
in the Antennae and evolves down and right to the area above the UCDs
and close to $\omega$~Cen. 

\section{Discussion}
\label{sec:disc}

With our numerical model we introduce a formation scenario for the
ultra-massive 'star cluster' W3 found in NGC~7252.  We propose that
this object may not have formed as one star cluster, but rather as a
star cluster complex.  The star clusters in this complex have merged
and formed W3.  These star cluster complexes (super-clusters) are a
common feature in interacting late-type galaxies.

We predict how W3 should look like if it would be possible to resolve
it.  Its age places it at about the transition time between having
many remaining star clusters visible inside and the time when all star 
clusters have already merged.  One should see the last surviving star
clusters in the stage of merging.  This should be also visible as
ripples in the surface brightness profile.

Furthermore a merger object as massive as W3 should show a distinct
core--envelope structure.  The merger object in our calculation had a
dense, dynamically cold core and an extended envelope.  This
core--envelope structure is not a transient feature but a stable
configuration.  We followed the evolution of our merger object for
$5$~Gyr and this structure remained.  The core is formed out of the
merged star clusters.  The envelope consists out of the stripped
stars, lost by the star clusters during the initial violent merger
process.  In our previous models the merger objects are not massive
enough to keep stripped stars bound and they are lost.  In the
case of the W3 model the core object is massive enough to retain the
stripped stars bound, forming the envelope structure of our model.

W3 is said to be one of the newly discovered ultra-compact dwarf (UCD) 
galaxies and also one of the most massive ones (Maraston et al.\
2004).  Other UCDs are found and studied around the central galaxy in
the Fornax cluster (NGC~1399) by Hilker et al.\ (1999), Phillipps et
al.\ (2001) and Mieske et al.\ (2004).  There these objects are quite
old.  There are two competing formation scenarios for UCDs.  While
Bekki et al.\ (2003) propose that these objects are the remaining
cores of stripped nucleated dwarf galaxies, Fellhauer \& Kroupa
(2002a) proposed the merging scenario from star cluster complexes as a
possible formation process, resulting naturally from the merging of
gas-rich galaxies in groups as the groups merge with a galaxy cluster.
A possible way to distinguish between the two scenarios would be an
analysis of the stellar populations in these objects.  While the
merging star cluster scenario implies that the stars of the UCDs have
more or less the same metalicity and age (this holds at least for the
most prominent population formed in the starburst), cores of dwarf
galaxies should show a more complex metalicity and age distribution.
And as dwarf galaxies are believed to be the oldest building blocks in
the universe, the cores should show a prominent very old population.
Differentiating between the two scenarios, however, will not be easy
because massive merger objects are likely to retain stellar winds,
thereby progressively building up an interstellar medium from which
new stars may form. They may also accrete interstellar gas and old
field stars from their progenitor galaxy (Kroupa 1998).

In the case of W3 the stripping scenario can be completely ruled out.
Age estimates of this object range from $300$ to $500$~Myr, which is
also the age of the interaction (NGC~7252 is a merger remnant of two
gas-rich disc galaxies).  This time is much too short for a nucleus to
be stripped of its surrounding dwarf galaxy.  If this implies that all
UCDs must have formed in the same way or if both scenarios are
possible and are realized by nature can not yet be asserted. 

Our model is an alternative to the formation of one single
massive object, which is believed to be highly unlikely even in
extreme starbursts as seen in interacting gas-rich galaxies.

\noindent {\bf Acknowledgements:}

\noindent MF thankfully announces financial support through DFG-grant 
KR1635/5-1.

\end{document}